\documentclass[aps,pra,preprint,superscriptaddress]{revtex4-2}

\usepackage[utf8]{inputenc}
\usepackage[english]{babel}
\usepackage[T1]{fontenc}

\usepackage{subfig}
\usepackage{bbold}
\usepackage{mathtools}
\usepackage{physics}

\newcommand{\one}{\mathbb{1}}

\DeclareMathOperator\sgn{sgn}

\begin{document}


\title{Self-testing of semisymmetric informationally complete measurements in a qubit prepare-and-measure scenario}


\author{Gábor Drótos}
\email[]{drotos.gabor@atomki.hu}
\affiliation{MTA Atomki Lend\"ulet Quantum Correlations Research Group, HUN-REN Institute for Nuclear Research, P.O. Box 51, H-4001 Debrecen, Hungary}
\author{Károly F. Pál}
\affiliation{HUN-REN Institute for Nuclear Research, P.O. Box 51, H-4001 Debrecen, Hungary}
\author{Tamás Vértesi}
\affiliation{MTA Atomki Lend\"ulet Quantum Correlations Research Group, HUN-REN Institute for Nuclear Research, P.O. Box 51, H-4001 Debrecen, Hungary}


\date{\today}

\begin{abstract}
Self-testing is a powerful method for certifying quantum systems. Initially proposed in the device-independent (DI) setting, self-testing has since been relaxed to the semi-device-independent (semi-DI) setting. In this study, we focus on the self-testing of a specific type of non-projective qubit measurements belonging to a one-parameter family, using the semi-DI prepare-and-measure (PM) scenario. Remarkably, we identify the simplest PM scenario discovered so far, involving only four preparations and four measurements, for self-testing the fourth measurement. This particular measurement is a four-outcome non-projective positive operator-valued measure (POVM) and falls in the class of semisymmetric informationally complete (semi-SIC) POVMs introduced by Geng et al. [Phys. Rev. Lett. 126, 100401 (2021)]. To achieve this, we develop analytic techniques for semi-DI self-testing in the PM scenario. Our results shall pave the way towards self-testing any extremal qubit POVM within a potentially minimal PM scenario.
\end{abstract}


\maketitle

\section{\label{sec:intro}Introduction}

\textit{POVMs in quantum theory.}---Measurement lies at the core of all physical sciences. In quantum mechanics, John von Neumann pioneered the concept of measurement by representing a projective measurement on a physical system by a complete set of orthogonal projectors on a Hilbert space~\cite{holevo2003}. However, in modern quantum theory, a more general measure called positive operator valued measure (POVM) describes the measurement~\cite{peres1997book,nielsen2002}. A POVM defines a set of positive operators ($M_i\succeq 0$) that sum up to the identity ($\sum_i M_i=\one$), without the constraint of $M_i^2=M_i$ as in the projective case. Non-projective POVMs offer advantages in quantum information, enhancing entanglement detection~\cite{shang2018,bae2019} and state discrimination~\cite{dieks1988,peres1988}. They find useful applications in various quantum protocols such as quantum coin flipping, quantum money, and quantum cryptography~\cite{bennett1992}. 

A paradigmatic example of a non-projective POVM is known as a symmetric informationally complete (SIC) POVM, which comprises rank-one POVM elements~\cite{renes2004}. A $d$-dimensional SIC can be interpreted both as a set of $d^2$ pure states $\{\ket{\psi_i}\}_i$ and as a POVM $\{M_i\}_i$ with $d^2$ elements $M_i=(1/d)\ket{\psi_i}\bra{\psi_i}$. Each element corresponds to a distinct outcome, and they satisfy $\Tr(M_i)=1/d$ for all $i$. Additionally, the elements of the POVM $\{M_i\}_i$ satisfy the condition
\begin{equation}
    \Tr(M_iM_j)=
    \frac{\left|\bra{\psi_i}\ket{\psi_j}\right|^2}{d^2}=\frac{1}{d^2(d+1)}
    \label{rhoihoiv}
\end{equation}
for all $i\neq j$. Indeed, this condition ensures the completeness of the POVM, i.e., $\sum_i M_i=\one$. While the simplest explicit construction of a SIC POVM is in dimension two, it can be constructed in higher dimensions as well. In particular, it is conjectured that SIC POVMs exist in all finite dimensions $d\ge 2$~\cite{zauner2011}, although a general proof is yet to be established. A comprehensive review of SIC POVMs can be found in Ref.~\cite{fuchs2017}. In fact, SIC POVMs belong to the broader class of informationally complete (IC) measurements~\cite{dariano2004}. This class has the useful property that the probabilities for the outcomes of any measurement on a target system can be computed if the probability distribution for the outcomes of an IC measurement on the same target system is known.

Within mathematics, SIC POVMs have deep connections with prominent open problems in algebraic number theory, including Hilbert's 12th problem~\cite{appleby2017}. On the other hand, their significance in physics arises from their key role in quantum information theory. They are exploited in a number of protocols such as optimal state tomography~\cite{wootters89,scott2006}, quantum key distribution~\cite{fuchs2003,renes2005}, entanglement detection~\cite{shang2018}, dimension witnessing~\cite{brunner2013dimension}, device-independent randomness generation~\cite{acin2016optimal,andersson2018}, and certification of measurement devices~\cite{tavakoli2020self,mironowicz2019}.

Recently, Geng et al.~\cite{Geng2021} introduced a broader class of POVMs, called semisymmetric informationally complete (semi-SIC) POVMs, which extends beyond SIC POVMs. Semi-SIC POVMs are derived from SIC POVMs by dropping the condition $\Tr(M_i)=1/d$ for all outcomes $i=(1,\ldots,d^2)$ while prescribing the conditions 
\begin{equation}
    \Tr(M_iM_j)=B
    \label{semicondi}
\end{equation}
on the elements of $\{M_i\}_i$, which are still required to be rank-one, for all $i\neq j$; it turns out that $B \in (1/16,1/12]$ for $d=2$ and can take only a few discrete values for $d\geq 3$.

It should be noted that other generalisations of SIC POVMs are also possible. For example, Appleby~\cite{appleby2007symmetric} discusses the class of IC POVMs that are symmetric, but the elements $M_i$ are not necessarily rank-one. 

In this paper, our focus is on self-testing measurements from the class of semi-SIC POVMs in dimension two within the so-called prepare-and-measure (PM) protocol~\cite{gallego2010device}. In the context of self-testing, we will briefly introduce the most stringent device-independent (DI) scenario, and subsequently address the semi-device-independent (semi-DI) PM scenario, which will be the subject of our analysis.

\textit{Self-testing of quantum systems.}---Self-testing of quantum systems is a recent approach in quantum information, which is partly related to security and robustness issues. It attempts to formulate protocols and experiments in a device-independent (DI) way~\cite{acin2007device,scarani2012device}. The concept of self-testing within the DI scenario was originally proposed by Mayers and Yao~\cite{mayers2004}; it means that the characterisation of the preparation and measurement apparatuses are solely based on the observed measurement statistics. Importantly, no assumptions are made about the internal functioning of the devices, made possible by exploiting the Bell nonlocal property of multipartite quantum correlations~\cite{bell1964,brunner2014bell}. 

A popular relaxation of the DI approach is the so-called semi-device-independent (semi-DI) setting, where certain additional assumptions are made about the devices~\cite{pawlowski2011semi}. These physical assumptions may involve bounds on the overlap~\cite{brask2017megahertz}, mean energy~\cite{van2017semi}, entropy~\cite{chaves2015device} of the prepared states, or may be based on the response of the physical systems to spatial rotations~\cite{jones2022}. In the standard PM scenario, however, it is customary to assume an upper bound on the dimension of the communicated quantum system~\cite{tavakoli2018self,farkas2019self,mironowicz2019,tavakoli2020self,divianszky2022, navascues2023}.

Let us now summarise the self-testing results that have been achieved in the literature within a PM scenario. Essentially, three main tasks have been demonstrated: self-testing of (i) quantum states, (ii) projective measurements, and (iii) non-projective POVM measurements. It is worth noting that most of these results are related to the $n\rightarrow 1$ quantum random access code (QRAC) scenario~\cite{nayak1999,ambainis2002,ambainis2008quantum}. This scenario can be considered as a special case of the PM scenario, where the goal is to encode $n$ digits into a single message qudit, and the receiver attempts to extract one of the digits. 

In the case of self-testing of quantum states, a specific set of four qubit states, namely $\{\ket{0},\ket{1},\ket{+},\ket{-}\}$  has been shown to be self-testable within the $2\rightarrow 1$ QRAC scenario~\cite{tavakoli2018self}. Furthermore, for the $3\rightarrow 1$ QRAC, it has been proven that the self-testing conditions can certify that the eight prepared states correspond to Bloch vectors forming a cube on the Bloch sphere. 

Regarding the self-testing of projective measurements, the biased QRAC scenario in Refs.~\cite{tavakoli2018self,alves2023biased} could self-test any pair of incompatible qubit projective measurements. In fact, these references could even further self-test certain triples of qubit projective measurements, including those corresponding to three mutually unbiased bases~\cite{schwinger1960,bengtsson2007}. Ref.~\cite{tavakoli2018self} uses the $2\rightarrow 1$ QRAC protocol to self-test a non-trivial pair of qutrit measurements. These results have subsequently been extended to higher dimensional systems in a more didactic manner. Specifically, Ref.~\cite{farkas2019self} uses QRAC with a higher-dimensional message to self-test a pair of $d$-dimensional measurements based on mutually unbiased bases for arbitrary $d$ (see also \cite{alves2023biased}). 

In the case of self-testing non-projective measurements, references~\cite{mironowicz2019,tavakoli2020self} achieve self-testing of an extremal four-outcome qubit POVM known as the qubit SIC POVM. To this end, Ref.~\cite{mironowicz2019} uses the $3\rightarrow 1$ QRAC, while Ref.~\cite{tavakoli2020self} employs a truncated version of the $3\rightarrow 1$ QRAC. Additionally, Ref.~\cite{tavakoli2019enabling} robustly self-tests $d$-dimensional SIC POVMs, and Ref.~\cite{Tavakoli2020b} self-tests measurements compounded by multiple SICs. Furthermore, Pauwels et al.~\cite{pauwels2022} recently certified a real-valued qubit POVM in a minimal scenario. However, this task is weaker than self-testing since it does not establish a connection to a target POVM. Let us also mention a more recent work which provides a family of linear witnesses whose maximum value self-tests arbitrary ensembles of pure states and arbitrary sets of extremal POVMs in PM scenarios of arbitrary Hilbert space dimension~\cite{navascues2023}. It should be noted, however, that the number of preparations and measurements in this work does not in general define a minimal PM scenario. In the meantime, we have also published general and possibly minimal constructions for self-testing in qubit Hilbert spaces~\cite{drotos2024}, for which certain proofs are provided in the present paper, which we had compiled and made accessible earlier; note also that the witness constructed for self-testing semi-SIC POVMs in our present work does not match any of the general constructions.

\textit{Organisation of the paper.}---In this research, we present novel analytic methods for self-testing members of the one-parameter family of non-projective qubit measurements introduced by Geng et al.~\cite{Geng2021}, the qubit semi-SIC POVMs, in a semi-DI manner. To this end, we focus on a PM scenario, assuming a two-dimensional bound on the Hilbert space. Firstly, in Section~\ref{sec:semiSICpovm}, we define the qubit semi-SIC POVMs, characterised by the continuous parameter $B$; we follow a formulation equivalent to that of Geng et al.~\cite{Geng2021}. Moving on, in Section~\ref{sec:selftestPM}, we define the general framework for self-testing quantum states and measurements within the qubit PM scenario. The specific setup we employ to self-test non-projective measurements is adopted from Ref.~\cite{tavakoli2020self} and consists of two stages which are applied in our study as follows.

(i) In Section~\ref{sec:states}, we construct a two-parameter ($c_1,c_2$) witness matrix $w$ with the purpose of self-testing four specific qubit states, parameterised by $c_1$ and $c_2$. These states are shown to be self-tested by the maximum value of the witness if their Bloch vectors correspond to those associated with a semi-SIC POVM, ensured by an appropriate choice of $c_1$ and $c_2$ as a function of the parameter $B$ of the semi-SIC POVM, as described in Section~\ref{sec:correspondence}.

(ii) Next, in Section~\ref{sec:semi-SIC}, it is shown that we can achieve self-testing of our particular semi-SIC POVM using the self-tested states mentioned above. Importantly, all of our results are derived analytically and applicable in the noiseless scenario. 

Finally, we conclude the paper with a discussion in Section~\ref{sec:disc}, where we also present open problems related to the minimal configuration and the noisy version of the PM scenario. Our analytic approach shall serve as a stepping stone towards self-testing any extremal qubit POVM in a minimal PM scenario.

\section{\label{sec:semiSICpovm}The qubit semi-SIC POVM}

In this section, we will discuss the qubit semi-SIC POVM and its representation in terms of Bloch vectors. Note that if $\{E_i\}_{i=1}^o$ defines a qubit POVM with $o$ outcomes (that is $\sum_i^o E_i=\one_2$ and $E_i\succeq 0$; in what follows, we shall omit the index $2$ for $\one$), then it can be expressed as:
\begin{equation}
\label{Eiext}
E_i=a_i(\one+\vec h_i\cdot\vec\sigma)
\end{equation}
for $i=(1,\ldots,o)$, where $|\vec h_i|\le 1$, $a_i>0$, and $\vec\sigma=(\sigma_x,\sigma_y,\sigma_z)$ is the vector of Pauli matrices. The condition $\sum_{i=1}^o E_i=\one$ is equivalent to $\sum_{i=1}^o a_i=1$ and $\sum_{i=1}^o a_i\vec h_i=0$.

We can consider a POVM equivalent to another POVM up to certain unitary or antiunitary transformations as follows.
If $\{E_i\}_i$ defines a qubit POVM, than the elements $\{UE_iU^{\dagger}\}_i$ also define a valid qubit POVM, where $U$ is an arbitrary $2\times 2$ unitary or anti-unitary operator. The following set of POVMs is equivalent to the semi-SIC POVMs constructed in Ref.~\cite{Geng2021} and given by Eqs.~(9-10) in that reference. We have chosen the form below, because the symmetries are more transparent in this representation. Namely, our semi-SIC POVM has the following elements:  
\begin{align}
 E_1&=\frac{a_-}{2}\left(\one+\vec h_1\cdot\vec\sigma\right),\nonumber\\
 E_2&=\frac{a_-}{2}\left(\one+\vec h_2\cdot\vec\sigma\right),\nonumber\\
 E_3&=\frac{a_+}{2}\left(\one+\vec h_3\cdot\vec\sigma\right),\nonumber\\
 E_4&=\frac{a_+}{2}\left(\one+\vec h_4\cdot\vec\sigma\right),
 \label{semiEi}
\end{align}
where $a_{\pm}=(1\pm\sqrt{1-12B})/2$, and we choose the unit vectors $\vec h_i$ as follows:
\begin{align}
\label{h1to4}
 \vec h_1&=\left(r_-,r_-,q_-\right), &\vec h_3&=\left(-r_+,r_+,-q_+\right),\nonumber\\
 \vec h_2&=\left(-r_-,-r_-,q_-\right),&\vec h_4&=\left(r_+,-r_+,-q_+\right),
\end{align}
where $r_{\pm}=\sqrt{(1-q_{\pm}^2)/2}$ and $q_{\pm}=\sqrt B/a_{\pm}$ (due to normalisation). It can be readily checked that 
$\Tr(E_iE_j)=B$ for all $i\neq j$, and $\Tr(E_1)=\Tr(E_2)=a_-$, $\Tr(E_3)=\Tr(E_4)=a_+$. On the other hand, 
\begin{equation}
\frac{a_-(\vec h_1+\vec h_2)+a_+(\vec h_3 + \vec h_4)}{2}=a_-q_--a_+q_+=0,
\end{equation}
hence $\{E_i\}_i$ indeed define a valid POVM with parameter $B$. This is the so-called semi-SIC POVM. The allowed range of the parameter $B$ is $B\in\left(1/16,1/12\right]$. For $B=1/16$ the first two elements are $E_1=E_2$, and therefore the POVM is not informationally complete. Note that for $B=1/12$ we have $a_{\pm}=1/2$ and $q_{\pm}=1/\sqrt 3=r_{\pm}$, that is the four $\vec h_i$ unit vectors define the four corners of a regular tetrahedron, which corresponds to the qubit SIC POVM. 

\begin{figure}
  \subfloat{\label{fig:dd_b_1p15}\includegraphics[trim={2.5cm 0 2.5cm 0},clip,width=0.3\textwidth]{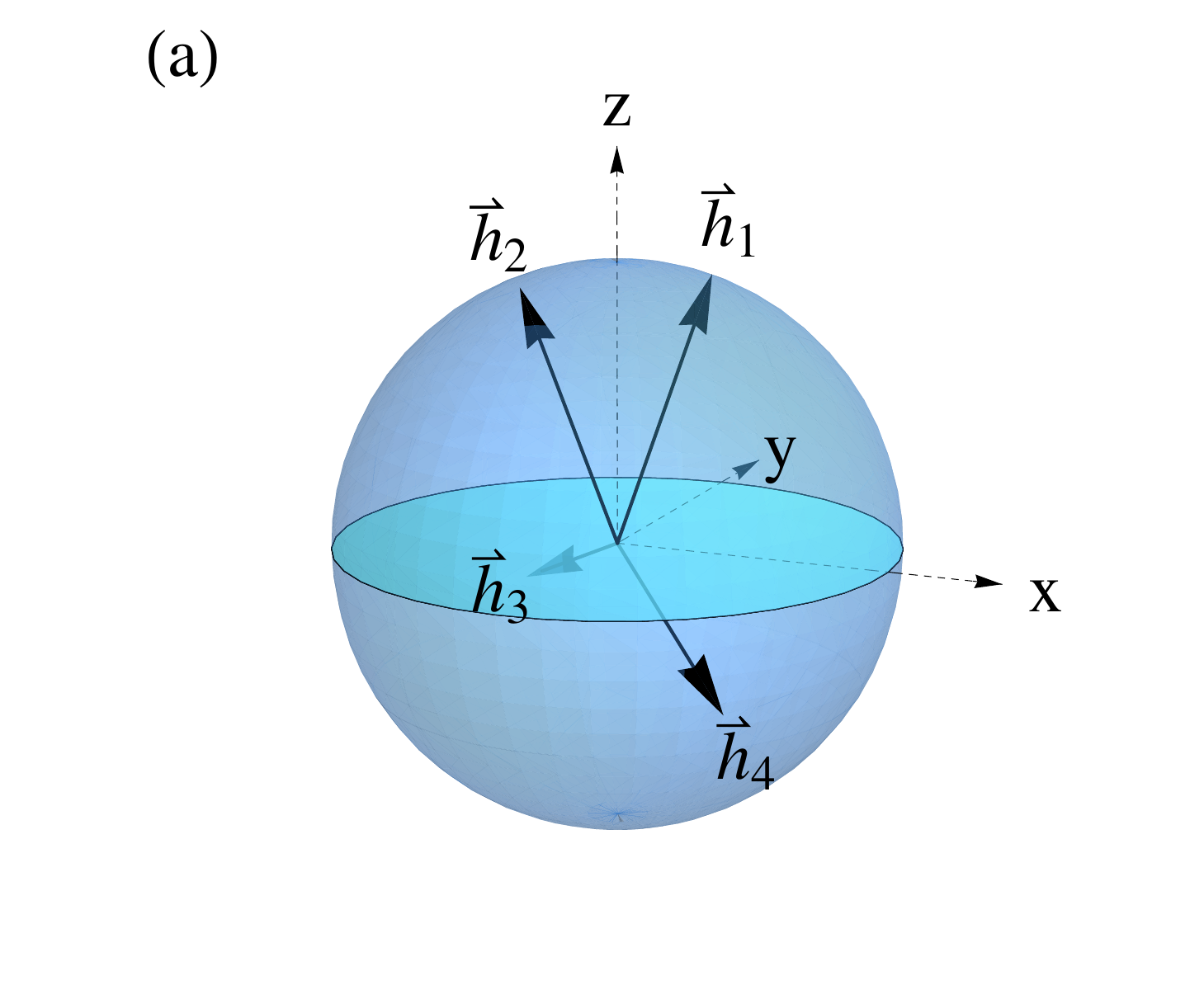}}
  \subfloat{\label{fig:dd_b_1p12}\includegraphics[trim={2.5cm 0 2.5cm 0},clip,width=0.3\textwidth]{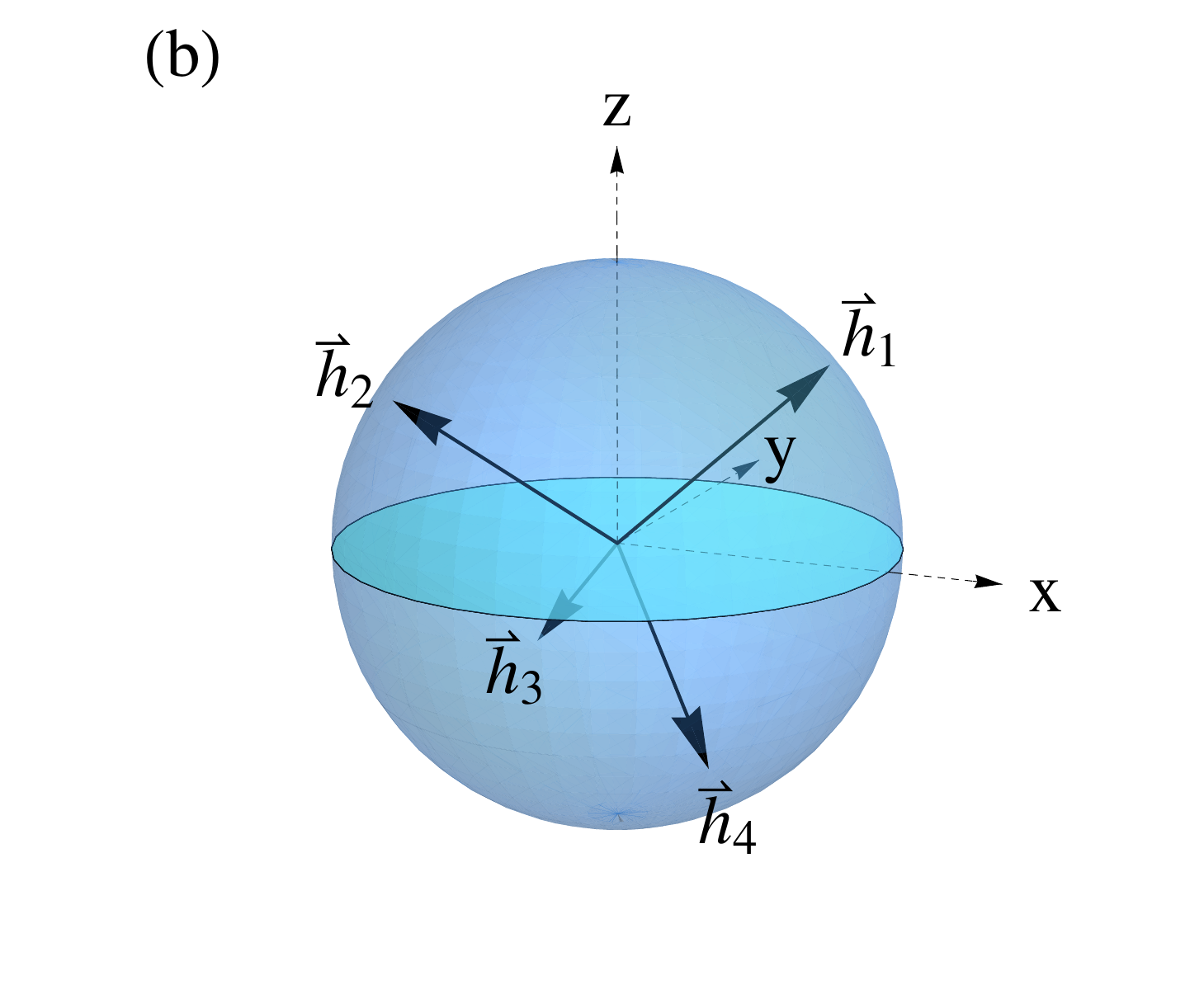}}
\caption{\label{fig:dd}The Bloch vectors \eqref{h1to4} of a semi-SIC POVM for (a) $B = 1/15$ and (b) $B = 1/12$. Note that panel (b) corresponds to a SIC POVM.}
\end{figure}

Note that we have the following six edges $\vec h_{i,j}=\vec h_i-\vec h_j$ from the $\vec h_i$ vectors in \eqref{h1to4}: 
\begin{align}
 \vec h_{12} &= (2r_-,2r_-,0),\nonumber\\
 \vec h_{13} &= (r_-+r_+,r_--r_+,q_-+q_+),\nonumber\\
 \vec h_{14} &= (r_--r_+,r_-+r_+,q_-+q_+),\nonumber\\
 \vec h_{23} &= (r_+-r_-,r_++r_-,q_-+q_+),\nonumber\\
 \vec h_{24} &= (-r_--r_+,-r_-+r_+,q_-+q_+),\nonumber\\
 \vec h_{34} &= (-2r_+,2r_+,0),
\end{align}
where we can see that the length of the four edges $\vec h_{13}, \vec h_{14}, \vec h_{23}$ and $\vec h_{24}$ is the same. The other two ($\vec h_{12}$ and $\vec h_{34}$) are orthogonal to each other. This object is called a digonal disphenoid~\cite{DD}. The Bloch vectors spanning this digonal disphenoid are illustrated in Fig.~\ref{fig:dd} for $B = 1/15$ and $B = 1/12$. In the latter case, corresponding to a SIC POVM, the digonal disphenoid is a regular tetrahedron.

We call a POVM extremal if it cannot be decomposed as a convex mixture of other POVMs. Extremal qubit POVMs can have $o=2,3$ or $4$ outcomes. In the case of four outcomes ($o=4$), an extremal qubit POVM must have four unit vectors $\vec h_i$ of which no choice of three lies in the same plane~\cite{dariano2005}. Only extremal POVMs can be self-tested, as the statistics of nonextremal POVMs can be simulated by stochastically implementing extremal POVMs (more on this in the next section). For the parameter $B$ in the range $B\in\left(1/16,1/12\right]$, the unit vectors in \eqref{h1to4} defining the vertices of the digonal disphenoid are such vectors, making qubit semi-SIC POVMs possible candidates for self-testing.

\section{\label{sec:selftestPM}Self-testing in the qubit PM scenario}

In this section, we will define the concept of self-testing quantum states and measurements in the qubit PM scenario. It is important to note that the PM scenario differs from Bell scenarios in two key points. First, in the PM scenario, the parties are communicating and thus there is no space-like separation between them. Second, unlike a Bell scenario, the PM scenario does not involve entanglement. The PM scenario can be modelled by two separated parties, say Alice and Bob, as follows (see the scenario shown in Fig.~\ref{fig:pm}).

Upon receiving an input $x \in \{1,\ldots,m\}$, Alice prepares a state $\rho_x$ and sends it to Bob. Bob also receives an input $y \in \{1,\ldots,n\}$ and performs a measurement depending on $y$ described by the operators $\{M_{b|y}\}_{by}$, where $b\in\{0,\ldots,o-1\}$ or $b\in\{1,\ldots,o\}$ denotes the outcome of the measurement (the choice for indexing may depend on $y$). It is worth mentioning that the value of $o$ may also depend on $y$. This setup generates a probability distribution given by
\begin{equation}\label{eq:probbxy}
	\mathcal{P}(b|x,y) = \Tr(\rho_x M_{b|y}).
\end{equation}

The goal of self-testing is to characterise a quantum system based only on the observed data, i.e., represented by \eqref{eq:probbxy}. Self-testing means to infer from the above statistics that the preparations and measurements, denoted by $\{\rho_x,M_{b|y}\}$, are equivalent to some reference or target set $\{\rho_x',M_{b|y}'\}$ up to a known transformation (referred to as an isometry).

A self-test of the target set can be achieved using a witness $\mathcal{W}$, which is a linear function of the probability distribution $\mathcal{P}(b|x,y)$:
\begin{equation}\label{eq:Wdef}
	\mathcal{W} = \sum_{x,y,b} w_{xy}^{(b)} \mathcal{P}(b|x,y).
\end{equation}
Our objective is to construct a witness $\mathcal{W}$ such that its maximum value $Q$ attainable with qubit systems self-tests the target states and measurements $\{\rho_x',M_{b|y}'\}$.

Note that we can self-test only pure states (i.e., $\rho'_x=\ket{\psi'_x}\bra{\psi'_x}$) and extremal POVMs. This limitation arises from the fact that the maximum value of the linear witness $\mathcal{W}$ for a given dimension $d$ can always be achieved with pure states and extremal measurements (see the proof of Lemma~1 in Ref.~\cite{ahrens2014} regarding the purity of the states; the proof is analogous for the extremality of the measurements). Considering that $\mathcal{P}(b|x,y)=\mathcal{P}(b|x,y)^*$, we have $\mathcal{P}(b|x,y)=\Tr(\rho_x M_{b|y})=\Tr(\rho_x^* M_{b|y}^*)$. Additionally, since a self-tested state must be pure, any probability $\mathcal{P}(b|x,y)$ obtained using $\left\{\ket{\psi_x},M_{b|y}\right\}$ can also be obtained with either $\left\{\ket{\psi_x}^*,M^*_{b|y}\right\}$ or $\left\{U\ket{\psi_x},UM_{b|y}U^{\dagger}\right\}$, where $U$ is an arbitrary $2\times 2$ unitary. Note that with an important additional assumption on the preparations and measurements that they are independent and identically distributed (i.i.d.~case), Ref.~\cite{miklin2021} could show that even mixed qubit states can be self-tested, which is an impossible task in our non-i.i.d. case.

Let us now examine how the above statements can be translated to the Bloch representation of states and measurements. Namely, we represent qubit states in the Bloch form as follows:
\begin{equation}\label{eq:qubit}
\rho_x=\frac{\one+\vec m_x\cdot\vec\sigma}{2},
\end{equation}
where the Bloch vector $\vec m_x$ has a unit length $|\vec m_x|=1$, indicating a pure state. We then consider the Bloch vectors $\vec h_b$ of a generic four-outcome qubit POVM defined by \eqref{Eiext}. In this case, we have the following relationships between the specific vectors ($\vec m_x$, $\vec h_b$) and the target vectors ($\vec m'_x$, $\vec h'_b$):
\begin{align}
\vec m_x=R\vec m'_x,\;\; \vec h_b=R\vec h'_b
\end{align}
or
\begin{align}
\vec m_x=-R\vec m'_x, \;\; \vec h_b=-R\vec h'_b 
\end{align}
for all $x$ and $b=(1,\ldots,4)$, where $R$ is an arbitrary three-dimensional rotation matrix.  

The objective of this paper is to prove, up to an isometry $\pm R$, the equivalence between the Bloch vectors of the target POVM $\{M'_{b|y}\}$ and those of the semi-SIC POVM defined by \eqref{semiEi} when the witness \eqref{eq:Wdef} achieves its maximum value $Q$ during the PM protocol.

\section{\label{sec:states}Self-testing specific qubit states in the PM scenario}

\begin{figure}
  \subfloat{\label{fig:pm_basic}\includegraphics[width=0.375\textwidth]{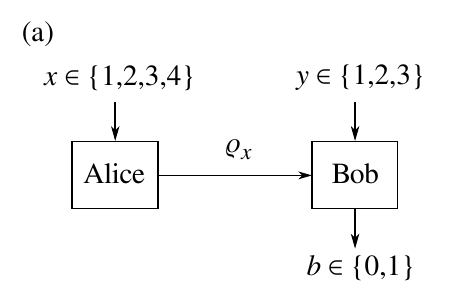}}
  \subfloat{\label{fig:pm_extension}\includegraphics[width=0.375\textwidth]{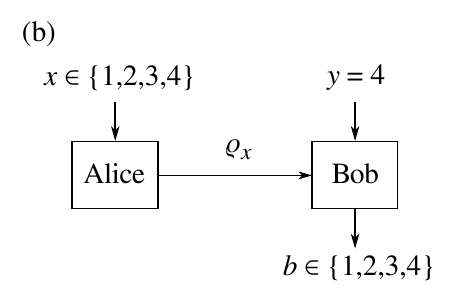}}
  \caption{\label{fig:pm}The prepare-and-measure scenario considered in this study. Panel (a) shows the setup corresponding to the first part of the self-testing problem, while panel (b) shows its extension.}
\end{figure}

Below we discuss in detail a PM experiment, which involves two-dimensional (i.e., qubit) systems. In the first part of the self-testing problem, our goal is to self-test four specific pure qubit states. Accordingly, we will restrict our considerations to four preparations ($m = 4$), three measurement settings ($n = 3$) with two outcomes per setting ($o = 2$), and take $\rho_x$ to be a generic qubit. This implies a density operator of the form \eqref{eq:qubit} for the prepared states, and we write $M_{b|y}$ in the most general form of a two-outcome POVM element,
\begin{equation}\label{eq:measurement_general}
	M_{b|y} = \left(1+(-1)^b\mu_y\right) \frac{\one+(-1)^b\vec{v}_y^{(b)} \cdot \vec{\sigma}}{2} ;
\end{equation}
for the sake of completeness, we allow for any Bloch vectors $\vec{m}_x$ and $\vec{v}_y^{(b)}$ with lengths $|\vec{m}_x| \leq 1$ and $|\vec{v}_y^{(b)}| \leq 1$ and any parameters $\mu_y \in [-1,1]$ except that we require $(1+\mu_y)\vec{v}_y^{(0)} = (1-\mu_y)\vec{v}_y^{(1)}$ to be satisfied. See Fig.~\ref{fig:pm_basic} for this setup.

For the aim of self-testing, we define the witness according to \eqref{eq:Wdef} where, in order to be able to self-test semi-SIC POVMs later, we choose the matrices $w^{(b)}$ as
\begin{equation}\label{eq:w}
	w^{(0)} = -w^{(1)} = w =
	\begin{pmatrix*}[r]
		c_1 & 1 & 1 \\
		c_1 & -1 & -1 \\
		-c_2 & 1 & -1 \\
		-c_2 & -1 & 1
	\end{pmatrix*}
	,
\end{equation}
where $c_1$ and $c_2$ are arbitrary real parameters. Note that $\mathcal{W} = \sum_{x,y} w_{xy} E_{xy}$, where $E_{xy}$ is the expectation value of an observable that takes on the values $1$ and $-1$ upon the measurement outcomes $b = 0$ and $b = 1$, respectively.

The prepared states $\rho_x$ can only be self-tested if $\vec{m}_x$, $\vec{v}_y^{(b)}$ and $\mu_y$ maximise $\mathcal{W}$. We now assume $\mu_y = 0$ for all $y$, in which case \eqref{eq:measurement_general} simplifies to an operator describing a non-degenerate projective measurement:
\begin{equation}\label{eq:measurement_projective}
	M_{b|y} = \frac{\one+(-1)^b\vec{v}_y \cdot \vec{\sigma}}{2} ,
\end{equation}
where $\vec{v}_y = \vec{v}_y^{(0)} = \vec{v}_y^{(1)}$. From \eqref{eq:probbxy} with \eqref{eq:qubit} and \eqref{eq:measurement_projective}, it turns out that
\begin{equation}\label{eq:Wmv}
	\mathcal{W} = \sum_{x,y} w_{xy} \vec{m}_x \cdot \vec{v}_y .
\end{equation}

Let us define
\begin{equation}\label{eq:u}
	\vec{u}_x = \sum_y w_{xy} \vec{v}_y ,
\end{equation}
with which we can write
\begin{equation}
	\mathcal{W} = \sum_{x} \vec{m}_x \cdot \vec{u}_x .
\end{equation}
If $\vec{u}_x \neq 0$ for all $x$, then $\mathcal{W}$ is obviously maximised (taking into account that $|\vec{m}_x|$ is maximally $1$) by choosing
\begin{equation}\label{eq:mmax}
	\vec{m}_x = \frac{\vec{u}_{x}}{\left|\vec{u}_{x}\right|} ,
\end{equation}
from which
\begin{equation}\label{eq:Qdef}
	Q := \max_{\vec{m}_x,\vec{v}_y} \mathcal{W} = \max_{\vec{v}_y} \sum_{x} \left| \vec{u}_{x} \right| ,
\end{equation}
where the equality is true even if $\vec{u}_x = 0$ for some $x$, in which case $\vec{m}_x$ remains unspecified.

Let us introduce
\begin{equation}\label{eq:Qvdef}
	Q_v = \sum_x \left| \vec{u}_{x} \right| ,
\end{equation}
which is to be maximised in $\vec{v}_y$. In the main text, we follow a procedure that is straightforward to apply to more general witness matrices $w$; see Appendix~\ref{sec:means} for a simpler option specific to the form in \eqref{eq:w}. For the procedure of the main text, we demonstrate in Appendix~\ref{sec:configuration} that $Q_v$ cannot be maximal in $\vec{v}_y$ if $\vec{u}_x = 0$ for any $x$, so that \eqref{eq:mmax} holds for maximising $\mathcal{W}$: the optimal vectors $\vec{m}_x$ are unit vectors and are parallel with the vectors $\vec{u}_x$. It can be similarly shown that the length of the optimal vectors $|\vec{v}_y|$ must also be $1$. Then, by \eqref{eq:u} and \eqref{eq:w}, \eqref{eq:Qvdef} can be expanded as
\begin{widetext}
\begin{align}\label{eq:sumu}
	Q_v =& \sqrt{c_1^2+2+2(c_1\vec{v}_1\cdot\vec{v}_2+c_1\vec{v}_1\cdot\vec{v}_3+\vec{v}_2\cdot\vec{v}_3)} + \sqrt{c_1^2+2+2(-c_1\vec{v}_1\cdot\vec{v}_2-c_1\vec{v}_1\cdot\vec{v}_3+\vec{v}_2\cdot\vec{v}_3)} \nonumber \\
	&+ \sqrt{c_2^2+2+2(-c_2\vec{v}_1\cdot\vec{v}_2+c_2\vec{v}_1\cdot\vec{v}_3-\vec{v}_2\cdot\vec{v}_3)} + \sqrt{c_2^2+2+2(c_2\vec{v}_1\cdot\vec{v}_2-c_2\vec{v}_1\cdot\vec{v}_3-\vec{v}_2\cdot\vec{v}_3)} .
\end{align}
\end{widetext}
Let us define
\begin{align}
	\gamma_{12} &= \vec{v}_1\cdot\vec{v}_2 , \label{eq:gamma12} \\
	\gamma_{13} &= \vec{v}_1\cdot\vec{v}_3 , \label{eq:gamma13} \\
	\gamma_{23} &= \vec{v}_2\cdot\vec{v}_3 . \label{eq:gamma23}
\end{align}
The maximum of \eqref{eq:sumu} is found where its derivatives with respect to $\gamma_{12}$, $\gamma_{13}$ and $\gamma_{23}$ vanish; this is demonstrated in Appendix~\ref{sec:maximum}. Substituting \eqref{eq:u} after differentiation we obtain
\begin{align}
	\frac{c_1}{|\vec{u}_1|} - \frac{c_1}{|\vec{u}_2|} - \frac{c_2}{|\vec{u}_3|} + \frac{c_2}{|\vec{u}_4|} &= 0 , \label{eq:max1} \\
	\frac{c_1}{|\vec{u}_1|} - \frac{c_1}{|\vec{u}_2|} + \frac{c_2}{|\vec{u}_3|} - \frac{c_2}{|\vec{u}_4|} &= 0 , \label{eq:max2} \\
	\frac{1}{|\vec{u}_1|} + \frac{1}{|\vec{u}_2|} - \frac{1}{|\vec{u}_3|} - \frac{1}{|\vec{u}_4|} &= 0 . \label{eq:max3}
\end{align}
Adding and subtracting \eqref{eq:max1} and \eqref{eq:max2} yields
\begin{align}
	|\vec{u}_1| &= |\vec{u}_2| , \\
	|\vec{u}_3| &= |\vec{u}_4| ,
\end{align}
substituting which into $\eqref{eq:max3}$ results in
\begin{equation}
	|\vec{u}_3| = |\vec{u}_1| .
\end{equation}
That is,
\begin{equation}\label{eq:l}
	|\vec{u}_1| = |\vec{u}_2| = |\vec{u}_3| = |\vec{u}_4| =: l ,
\end{equation}
which, according to \eqref{eq:u} and \eqref{eq:gamma12}-\eqref{eq:gamma23}, means
\begin{align}
	l^2 &= c_1^2+2+2(c_1\gamma_{12}+c_1\gamma_{13}+\gamma_{23}) , \label{eq:u1max1} \\
	l^2 &= c_1^2+2+2(-c_1\gamma_{12}-c_1\gamma_{13}+\gamma_{23}) , \label{eq:u2max1} \\
	l^2 &= c_2^2+2+2(-c_2\gamma_{12}+c_2\gamma_{13}-\gamma_{23}) , \label{eq:u3max1} \\
	l^2 &= c_2^2+2+2(c_2\gamma_{12}-c_2\gamma_{13}-\gamma_{23}) . \label{eq:u4max1}
\end{align}
Subtracting \eqref{eq:u2max1} from \eqref{eq:u1max1} and \eqref{eq:u4max1} from \eqref{eq:u3max1} gives
\begin{align}
	\gamma_{12} &= \gamma_{13} , \\
	\gamma_{12} &= - \gamma_{13} ,
\end{align}
that is,
\begin{align}
	\vec{v}_1\cdot\vec{v}_2 &= 0 , \label{eq:condition1} \\
	\vec{v}_1\cdot\vec{v}_3 &= 0 . \label{eq:condition2}
\end{align}
With these results, \eqref{eq:u1max1}-\eqref{eq:u4max1} simplify to
\begin{align}
	l^2 &= c_1^2+2+2\gamma_{23} , \\
	l^2 &= c_2^2+2-2\gamma_{23} ,
\end{align}
from which
\begin{equation}
	\vec{v}_2\cdot\vec{v}_3 = \frac{c_2^2-c_1^2}{4} =: \cos(2\theta) . \label{eq:condition3}
\end{equation}
\eqref{eq:condition1}, \eqref{eq:condition2} and \eqref{eq:condition3} imply that, up to an isometry,
\begin{align}
	\vec{v}_1 &= (1, 0, 0) , \label{eq:v1max} \\
	\vec{v}_2 &= (0, \cos\theta, \sin\theta) , \label{eq:v2max} \\
	\vec{v}_3 &= (0, \cos\theta, \sin(-\theta)) , \label{eq:v3max}
\end{align}
from which, using \eqref{eq:u},
\begin{align}
	\vec{u}_1 &= (c_1, 2\cos\theta, 0) , \label{eq:u1max} \\
	\vec{u}_2 &= (c_1, -2\cos\theta, 0) , \label{eq:u2max} \\
	\vec{u}_3 &= (-c_2, 0, 2\sin\theta) , \label{eq:u3max} \\
	\vec{u}_4 &= (-c_2, 0, -2\sin\theta) , \label{eq:u4max}
\end{align}
and the corresponding $\vec{m}_x$ vectors are obtained from \eqref{eq:mmax}. Furthermore, substituting \eqref{eq:condition1}, \eqref{eq:condition2} and \eqref{eq:condition3} to \eqref{eq:sumu}, we obtain
\begin{equation}\label{eq:Q}
	Q = 2 \sqrt{2(c_1^2+c_2^2+4)} .
\end{equation}
Measuring this $\mathcal{W}$ value self-tests the states characterised by \eqref{eq:u1max}-\eqref{eq:u4max} (up to an isometry). Note that we have assumed $\mu_y = 0$ for all $y$ to come to this result; we show in Appendix~\ref{sec:excluding} that relaxing this assumption can only lead to smaller $\mathcal{W}$ values if $c_1$ and $c_2$ are linked to the parameter $B$ of a semi-SIC POVM as described in the next section.

\section{\label{sec:correspondence}Correspondence between the specific states and semi-SIC POVMs}

For self-testing semi-SIC POVMs, it will be useful to relate the Bloch vectors of the states prepared by Alice to those of a semi-SIC POVM through linking $c_1$ and $c_2$ to the parameter $B$ of a semi-SIC POVM. Note that the results in Section~\ref{sec:states} are more general, since they are valid for any states and measurements characterised by Bloch vectors with any real $c_1$ and $c_2$ in \eqref{eq:v1max}-\eqref{eq:u4max} such that the condition $\mu_y = 0$ for all $y$ maximises $\mathcal{W}$.

For establishing a relationship, what we require is that the dot products between the Bloch vectors $\vec{m}_x$ of Alice's prepared states as given by \eqref{eq:mmax} with \eqref{eq:u1max}-\eqref{eq:u4max} (determined up to an isometry) be equal to the dot products between the Bloch vectors $\vec{h}_i$, $i \in \{1,2,3,4\}$,
of the semi-SIC POVM with parameter $B$, given in \eqref{h1to4}:
\begin{equation}\label{eq:dotproduct}
	\vec{m}_i \cdot \vec{m}_j = \vec{h}_i \cdot \vec{h}_j
\end{equation}
for all $i,j \in {1,2,3,4}$. Since $|\vec{m}_i| = 1$ in the case in question and also $|\vec{h}_i| = 1$ for all $i \in \{1,2,3,4\}$, this requirement is equivalent to requiring the angles between the Bloch vectors to be identical.

The dot products between the different vectors $\vec{h}_i$ can be computed from \eqref{h1to4} and are given as
\begin{align}
	\vec{h}_1 \cdot \vec{h}_2 &= \frac{1-15B+\sqrt{1-12B}}{9B} , \\
	\vec{h}_1 \cdot \vec{h}_3 &= \vec{h}_1 \cdot \vec{h}_4 = \vec{h}_2 \cdot \vec{h}_3 = \vec{h}_2 \cdot \vec{h}_4 = -\frac{1}{3} , \\
	\vec{h}_3 \cdot \vec{h}_4 &= \frac{1-15B-\sqrt{1-12B}}{9B} .
\end{align}
Substituting these expressions into \eqref{eq:dotproduct} determines the values of $c_1$ and $c_2$ as a function of $B$ up to a sign such that $c_1$ and $c_2$ have the same sign:
\begin{align}
	c_1 &= \pm\sqrt{ \frac{2\left(1-6B+\sqrt{1-12B}\right)}{24B-1} } , \label{eq:c1_of_B} \\
	c_2 &= \pm\sqrt{ \frac{2\left(1-6B-\sqrt{1-12B}\right)}{24B-1} } . \label{eq:c2_of_B}
\end{align}
With this choice of $c_1$ and $c_2$, the redundant expressions evaluate to identity. We thus conclude that with these values of $c_1$ and $c_2$, the Bloch vectors $\vec{m}_x$ defined by \eqref{eq:mmax} with \eqref{eq:u1max}-\eqref{eq:u4max} (determined up to an isometry) also characterise a semi-SIC POVM. In the special case of a SIC POVM, corresponding to $B = 1/12$ \cite{Geng2021}, \eqref{eq:c1_of_B}-\eqref{eq:c2_of_B} give $c_1 = c_2 = \pm 1$. According to \eqref{eq:Q}, \eqref{eq:c1_of_B}-\eqref{eq:c2_of_B} also imply
\begin{equation}\label{eq:Q_of_B}
	Q = 24 \sqrt{\frac{B}{24B-1}} ,
\end{equation}
which becomes $Q = 4\sqrt{3}$ for a SIC POVM. We show in Appendix~\ref{sec:excluding} that $\mu_y \neq 0$ for any $y$ can only lead to smaller values of $\mathcal{W}$ than in \eqref{eq:Q_of_B}, so that $\mu_y = 0$ for all $y$ is a justified assumption for maximising $\mathcal{W}$.

\section{\label{sec:semi-SIC}Self-testing semi-SIC POVMs}

For the second part of the self-testing problem, we provide, as an extension, a \emph{fourth} option for Bob's input, upon which he performs a measurement described by a POVM consisting of elements $M_{b|y=4}$, $b \in {1,2,3,4}$, acting on the two-dimensional Hilbert space; see Fig.~\ref{fig:pm_extension}. The operator $M_{b|y=4}$ corresponding to each result $b$ is described by a Bloch vector $\vec{n}_b$ and a coefficient $\lambda_b$:
\begin{equation}\label{eq:Mby4}
	M_{b|y=4} = \lambda_b \left( \one+\vec{n}_b\cdot\vec{\sigma} \right) .
\end{equation}
Since the operators $\{M_{b|y=4}\}_b$ constitute a POVM, we have that
\begin{equation}\label{eq:sumMby4}
	\sum_{b=1}^4 M_{b|y=4} = \one ,
\end{equation}
from which the coefficients $\lambda_b$ uniquely follow once the vectors $\vec{n}_b$ are given and are not co-planar (taking into account that the Pauli matrices and the identity matrix are linearly independent). Therefore, the POVM in question is fully characterised by the Bloch vectors $\vec{n}_b$ unless they lie in the same plane, in which case the POVM cannot be self-tested.

We now require that the probability of measuring the $x$th result when Alice prepares the $x$th state be zero; according to Born's formula, this means
\begin{equation}
	\mathcal{P}(b=x|x,y=4) \equiv \Tr(\rho_x M_{x|y=4}) = 0 .
\end{equation}
Substituting \eqref{eq:qubit} and \eqref{eq:Mby4}, this condition can be re-written as 
\begin{equation}
	\lambda_{x} \left( 1+\vec{m}_x\cdot\vec{n}_{x} \right) = 0 ,
\end{equation}
from which, assuming naturally that $\lambda_{x} \neq 0$, it follows that
\begin{equation}\label{eq:nx}
	\vec{n}_{x} = - \vec{m}_x
\end{equation}
with $|\vec{n}_x| = |\vec{m}_x| = 1$.

What we obtained implies the following: if we extend the witness \eqref{eq:Wdef}, following Ref.~\cite{tavakoli2020self}, as
\begin{equation}\label{eq:Wp}
	\mathcal{W}' = \mathcal{W} - k \sum_{x=1}^4 \mathcal{P}(b=x|x,y=4)
\end{equation}
for some positive constant $k$ (which can be chosen arbitrarily and would be relevant in a noisy setting \cite{tavakoli2020self}), it can take the value $\mathcal{W}$ only if \eqref{eq:nx} holds. As a consequence, if the vectors $\vec{m}_x$ correspond to the Bloch vectors of a semi-SIC POVM (as defined by \eqref{eq:mmax} with \eqref{eq:u1max}-\eqref{eq:u4max} and \eqref{eq:c1_of_B}-\eqref{eq:c2_of_B}, up to an isometry; see Section~\ref{sec:correspondence}), then the operators $\{M_{b|y=4}\}_b$ must also constitute a semi-SIC POVM: in particular, the one with the opposite Bloch vectors. This semi-SIC POVM is thus self-tested by measuring the value given by \eqref{eq:Q_of_B} as $\mathcal{W}'$ as defined by \eqref{eq:Wp}.

\section{\label{sec:disc}Discussions}

In this study, our focus was on the self-testing of the members of a particular one-parameter family of four-outcome non-projective qubit measurements (the semi-SIC POVMs) in the PM scenario. This qubit PM scenario provides a natural framework for our investigation. Previous research has successfully self-tested a qubit SIC POVM~\cite{mironowicz2019, tavakoli2020self}, as well as higher-dimensional SIC POVMs~\cite{brunner2013dimension, tavakoli2019enabling}. More recently, there have been results in self-testing compounds of SIC POVMs~\cite{Tavakoli2020b} and arbitrary extremal POVMs of arbitrary Hilbert space dimension~\cite{navascues2023}. We note here that in the so-called one-sided DI framework~\cite{supic2016}, which utilises an alternative assumption within the semi-DI approach, results on the task of self-testing arbitrary extremal POVMs could also be obtained~\cite{sarkar2022,sarkar2023}. However, in that scenario, the resource used for self-testing is EPR-steering~\cite{Schrodinger1935,Wiseman2007}.

In our analysis, we focused on the noiseless setup and approached the self-testing problem in the PM scenario analytically. It would certainly be useful to develop a noise-robust methodology. Moreover, it would be valuable to explore self-testing of non-projective measurements under different assumptions than those considered in our work. See e.~g. Refs.~\cite{brask2017megahertz, van2017semi,chaves2015device} for imposing other assumptions on the mediated particles in the PM scenario.

As we conclude this paper, we would like to highlight some open problems that remain in the qubit case. First of all, it would be intriguing to develop a self-testing method in the PM setup for the most general class of extremal four-outcome qubit POVMs, using as few preparations as possible. In particular, can we self-test any extremal qubit POVM using only four preparations and four measurements? If so, is this the minimal setup, or could there exist smaller setups that can be used for self-testing? We do know that a lower bound is given by four preparations and two measurement settings, one with two outcomes and the other with four outcomes. Note that there is a strong hint that a real-valued three-outcome qubit POVM can be self-tested using only two measurement settings~\cite{pauwels2022}. Our expectation is that the answers to these open questions will lead to further developments in this area.

\appendix

\section{\label{sec:means}Finding the global maximum via the relationship between arithmetic and quadratic means}

In order to maximise \eqref{eq:Qvdef} in $\vec{v}_y$ (regardless of whether the vectors $\vec{m}_x$ are well defined or not), we shall make use of the relationship
\begin{equation}\label{eq:means}
	\sum_{i=1}^N p_i \leq \sqrt{N\left(\sum_{i=1}^N p_i^2\right)}
\end{equation}
between arithmetic and quadratic means of real numbers $p_i$, $i \in \{1,\ldots,N\}$, where equality holds if and only if $p_i = p_j$ for all $i$ and $j$. In particular, we substitute $p_i = \left| \vec{u}_{i} \right|$ for $i \in \{1,2,3,4\}$. It then turns out from \eqref{eq:Qvdef} with \eqref{eq:u} that
\begin{equation}\label{eq:Qv_ub}
	Q_v \leq 2 \sqrt{2\left[\left(c_1^2+c_2^2\right)\vec{v}_1^2+2\left(\vec{v}_2^2+\vec{v}_3^2\right)\right]} .
\end{equation}

The right-hand side of \eqref{eq:Qv_ub} is obviously maximised by $|\vec{v}_y| = 1$ for all $y$. This maximum can also be reached by the left-hand side by requiring $|\vec{v}_y| = 1$ as well for all $y$ and $\left| \vec{u}_{i} \right| = \left| \vec{u}_{j} \right|$ for all $i,j$. It is easy to see that the latter condition leads to \eqref{eq:condition1}-\eqref{eq:condition2} and \eqref{eq:condition3} of the main text; since these requirements allow reaching the maximum of an upper bound of $Q_v$, they also maximise $Q_v$ itself. The reason for which these requirements allow reaching the maximum of the upper bound in question, i.e., the expression on the right-hand side of \eqref{eq:Qv_ub}, is that this right-hand side is independent of the dot products of the vectors $\vec{v}_y$, which is a special property of the witness matrix $w$ in \eqref{eq:w}.

Note that \eqref{eq:condition1}-\eqref{eq:condition2} and \eqref{eq:condition3} lead to well-defined vectors $\vec{u}_x \neq 0$, \eqref{eq:u1max}-\eqref{eq:u4max} of the main text (up to an isometry), so that \eqref{eq:mmax} also holds. This implies that the vectors $\vec{m}_x$ maximising the witness $\mathcal{W}$ in \eqref{eq:Wmv} have unit length, i.e., they describe pure states.

\section{\label{sec:configuration}Proof about the configuration of the optimal vectors}

Let us consider the following expression:  
\begin{equation}
\mathcal{W}=\sum_{x,y}w_{xy}\vec m_x\cdot\vec v_y ,
\label{eq:Wmv_repeated}
\end{equation}
where $w_{xy}$ is now any matrix having at least one nonzero element in each of its rows and columns. We are interested in finding vectors $|\vec m_x|\leq 1$, $|\vec v_y|\leq 1$ maximising the value of $\mathcal{W}$.

First let us choose any set of vectors $\vec v_y$ satisfying the conditions $|\vec v_y|\leq 1$. Then from
\begin{equation}
\mathcal{W}=\sum_{x}\vec m_x\cdot\left(\sum_y w_{xy}\vec v_y\right)
\label{eq:Wmv_grouped}
\end{equation}
it is easy to see that for the chosen set of $\vec v_y$ the optimal $\vec m_x$ is parallel with $\sum_y w_{xy}\vec v_y$ and normalised (i.e., its length is maximal) whenever $|\sum_y w_{xy}\vec v_y|\neq 0$, otherwise $\vec m_x$ is undefined: $\mathcal{W}$ does not depend on it. It is easy to see that
\begin{equation}
\max_{\vec m_x} \mathcal{W} = \sum_x\left|\sum_y w_{xy}\vec v_y\right| =: Q_v
\label{eq:Qm}
\end{equation}
in either case.

Now let us suppose that we have chosen $\vec v_y$ such that at least one of the terms, say the $x'$th one, in the equation above is zero, i.e., $|\sum_y w_{x'y}\vec v_y|=0$. Let $w_{x'y'}$ be nonzero. Let us consider two modifications of $\vec v_y'$: we either add the infinitesimal vector $\vec\delta$ to it, or subtract $\vec\delta$ from it. If $|\vec v_y'|=1$, both modifications are only allowed if $\vec v_y'\cdot \vec\delta=0$, in which case they correspond to infinitesimal rotations; in other cases this restriction is not necessary. Due to either of the modifications the term $|\sum_y w_{x'y}\vec v_y|$ will grow: its value will become $|w_{x'y'}\delta|$. If there are other zero terms in \eqref{eq:Qm} they will either grow as well, or remain unchanged. The nonzero terms are analytic functions of the vectors $\vec v_y$, therefore their sum will either grow due to one of the two modifications (which correspond to opposite directions), or it will only change at second or higher order. Therefore, if \eqref{eq:Qm} contains a zero term, its value can always be increased by modifying a vector appropriately, so such a choice of the vectors $\vec v_y$ cannot correspond to a maximum of $Q_v$. We have thus proven that there is no zero term for the optimal choice of $\vec v_y$.

It follows from this conclusion that all of the vectors $\vec m_x$ are well-defined unit vectors if the vectors $\vec v_y$ are optimal. One can prove analogously that $\vec v_y$ are also well-defined unit vectors if $\vec m_x$ are optimal.

\section{\label{sec:maximum}Finding the global maximum by differentiation}

We are interested in finding unit vectors $\vec v_y$ maximising the following expression:
\begin{align}
Q_v&=\sum_{x=1}^{M_m}\left|\sum_{y=1}^{M_v} w_{xy}\vec v_y\right| \nonumber \\
&=\sum_{x=1}^{M_m}\sqrt{s_x+2\sum_{y=2}^{M_v}\sum_{y'=1}^{y-1}t_{xyy'}\gamma_{yy'}},
\label{eq:Qv_general}
\end{align}
where $s_x\equiv \sum_y w_{xy}^2$, $t_{xyy'}\equiv w_{xy}w_{xy'}$ and $\gamma_{yy'}\equiv\vec v_y\cdot\vec v_{y'}$ is an off-diagonal element of the Gram matrix of the set of $\vec v_y$ vectors. The diagonal elements are $\gamma(yy)=1$. The set of the optimal $\vec v_y$ vectors is never unique, they are determined only up to a global orthogonal transformation. It is their Gram matrix that may be uniquely determined. It is important to note that the method outlined here may only work if $M_v$, the number of vectors, is equal to the dimension of the vector space.

The function $Q_v$ as a function of $\gamma_{yy'}$ maps the set of $M_v\times M_v$ positive semidefinite real matrices whose diagonal elements are one onto a subset of the real numbers: the domain of the function is where $\gamma_{yy'}$ are elements of such a matrix. However, if we consider $\gamma_{yy'}$ just as real numbers ignoring their definition, the formula on the rightmost side of \eqref{eq:Qv_general} is a well-defined real function of the variables $\gamma_{yy'}$ wherever the expression under the square root is non-negative in each term. This region, which is called the natural or implied domain of the expression, is larger than the domain of $Q_v$. Let us refer to the function with the extended domain as $Q_{\gamma}$. The domain of $Q_{\gamma}$ is contiguous and is bounded by the hyperplanes where any of the terms in the sum over $x$ is zero (the expressions under the square roots are linear functions of the variables). The function $Q_{\gamma}$ is analytic everywhere inside this whole region (but not at its boundaries).

What we are going to prove is that if the partial derivatives of $Q_{\gamma}$ with respect to all $\gamma_{yy'}$ vanish somewhere, then $Q_{\gamma}$ will have its global maximum there. If the condition for the partial derivatives holds in a single point, then the solution is unique. If the condition is satisfied outside the domain of $Q_v$, then although the expression does have its global optimum there, it is not the global optimum of $Q_v$. It cannot be reached with any set of $\vec v_y$ vectors, so it does not correspond to the solution of the problem. However, if the condition holds at any point belonging to the domain, it is a solution even if the point is at the boundary of the domain, where the Gramian is zero and the $\vec v_y$ are linearly dependent, because that point is the global maximum even in the larger implied domain.

What we will prove now is that the Hessian of $Q_{\gamma}$ is negative semidefinite everywhere inside its domain. For the sake of simplicity, let us denote the pair of indices $yy'$ with a single index $\alpha$, where $1\leq\alpha\leq M_v(M_v-1)/2$. Then 
\begin{equation}
Q_{\gamma}=\sum_{x=1}^{M_m}\sqrt{s_x+2\sum_{\alpha=1}^{M_v(M_v-1)/2}t_{x\alpha}\gamma_{\alpha}}\equiv\sum_{x=1}^{M_m} l_x ,
\label{eq:Qv1}
\end{equation}
The $\alpha\beta$ element of the Hessian of $Q_v$ is:
\begin{align}
\frac{\partial^2 Q_{\gamma}}{\partial\gamma_{\alpha}\partial\gamma_{\beta}}&=-\sum_{x=1}^{M_m}\frac{t_{x\alpha}t_{x\beta}}{l_x^3}=-\sum_{x=1}^{M_m}\frac{t_{x\alpha}}{l_x^{3/2}}\frac{t_{x\beta}}{l_x^{3/2}}\nonumber\\
&=-\sum_{x=1}^{M_m}R^T_{\alpha x}R_{x\beta},
\label{eq:Hess}
\end{align}
where $R_{x\beta}\equiv t_{x\beta}/l_x^{3/2}$. As the Hessian is minus one times the product of the transpose of a matrix with the matrix itself, it is negative semidefinite. Consequently, the function is concave everywhere, so if all of its partial derivatives are zero somewhere, it has its global maximum there, indeed. However, as we have already mentioned, if this place happens to be outside the domain of $Q_v$, it does not correspond to the solution. It may also happen that $w_{xy}$ is such that there is no place at all where all partial derivatives are zero. In these cases, the maximum is somewhere at the boundary of the domain, and one has to find some alternative method to determine the optimal solution for $Q_v$.

The rank of the Hessian is equal to the rank of the matrix $R$. The rank of a matrix does not change if we multiply its rows by nonzero numbers. Therefore, this rank is the same as that of the matrix $t$. As the matrix $t$ is independent of the variables, the rank of the Hessian is the same everywhere within the domain of $Q_{\gamma}$. Specifically, the Hessian is negative definite either everywhere or nowhere. If the first derivatives are zero only at one point, the Hessian must be negative definite there, consequently everywhere. It is obvious that this cannot happen if $M_m$, the number of rows of the matrix $R$ (and $t$), is less than $M_v(M_v-1)/2$. Furthermore, if all partial derivatives are zero somewhere, the rows of matrix $t$ must be linearly dependent, as $\sum_x t_{x\alpha}/l_x=0$ must hold for all $\alpha$. In this case the rank of the matrix $t$ is at most $M_m-1$, therefore, the Hessian can only be negative definite, and the solution for the maximum be unique if $M_m\geq M_v(M_v-1)/2+1$. The same result has been derived in Ref.~\cite{Vertesi2009}. If the solution is not unique, $w_{xy}$ is inappropriate for self-testing.

\section{\label{sec:excluding}Excluding generic two-outcome POVMs as Bob's measurements}

We demonstrate in this appendix that choosing $\mu_y \neq 0$ in \eqref{eq:measurement_general} for any $y$ can only lead to smaller values of $\mathcal{W}$ as defined by \eqref{eq:Wdef} than \eqref{eq:Q_of_B} if $c_1$ and $c_2$ are chosen according to \eqref{eq:c1_of_B}-\eqref{eq:c2_of_B}, so that generic two-outcome POVMs can be excluded as Bob's measurements: choosing the latter to be projective and non-degenerate allows self-testing Alice's prepared states.

With \eqref{eq:measurement_general} and \eqref{eq:w}, and according to Born's rule, \eqref{eq:Wdef} can be written as
\begin{equation}\label{eq:W_general_transformed}
	\mathcal{W} = \sum_{x,y} w_{xy} \Tr\left( \rho_x M_y \right) ,
\end{equation}
where
\begin{equation}\label{eq:observable_general}
	M_y = M_{0|y} - M_{1|y} = \mu_y \one + \left(1-|\mu_y|\right) \vec{v}_y \cdot \vec{\sigma}
\end{equation}
with
\begin{equation}\label{eq:v}
	\vec{v}_y =
	\begin{cases}
		\vec{v}_y^{(0)} & \text{if $\mu_y \leq 0$,} \\
		\vec{v}_y^{(1)} & \text{if $\mu_y \geq 0$.}
	\end{cases}
\end{equation}
With \eqref{eq:qubit}, \eqref{eq:W_general_transformed} can be further transformed as
\begin{align}\label{eq:W_general_transformed2}
	&\mathcal{W} = \sum_{x,y} w_{xy} \left[ \mu_y + \left(1-|\mu_y|\right) \vec{v}_y \cdot \vec{m}_x \right] \nonumber \\
    &= \sum_{y} \left[ \mu_y W_y + \left(1-|\mu_y|\right) \vec{v}_y \cdot \left( \sum_x w_{xy} \vec{m}_x \right) \right] ,
\end{align}
where
\begin{equation}\label{eq:W}
	W_y = \sum_x w_{xy} .
\end{equation}
Similarly to the reasoning in Section~\ref{sec:states}, the maximum of \eqref{eq:W_general_transformed2} is obtained by setting
\begin{equation}\label{eq:vmax_general}
	\vec{v}_y = \frac{\vec{z}_y}{\left|\vec{z}_y\right|}
\end{equation}
with
\begin{equation}\label{eq:z}
	\vec{z}_y = \sum_x w_{xy} \vec{m}_x ,
\end{equation}
under the condition that $\vec{z}_y \neq 0$ for any $y$, and $\vec{v}_y$ can be chosen arbitrarily for maximising $\mathcal{W}$ if $\vec{z}_y = 0$ for some $y$. In either case,
\begin{equation}\label{eq:Wdef_general}
	\max_{\vec{m}_x,\vec{v}_y,\mu_y} \mathcal{W} = \max_{\vec{m}_x,\mu_y} \sum_{y} \left[ \mu_y W_y + \left(1-|\mu_y|\right) |\vec{z}_y| \right] .
\end{equation}

In case $\left|\vec{z}_y\right| > \left|W_y\right|$ for a given $y$ and for a given set of vectors $\vec{m}_x$, $\mu_y = 0$ will give the maximal $\mathcal{W}$ for that $y$ in \eqref{eq:Wdef_general}; similarly, if $\left|\vec{z}_y\right| < \left|W_y\right|$ (note that this may not occur if $W_y = 0$), then $\mu_y = \sgn W_y$ will correspond to the maximum. One can then consider these two options for $\mu_y$ for each $y$ for which $W_y \neq 0$, maximise $\mathcal{W}$ in $\vec{m}_x$ and $\vec{v}_y$ for each configuration of the options, and compare the resulting values of $\mathcal{W}$ for finding the maximal $\mathcal{W}$ in $\mu_y$. This is sufficient, because in case one takes an optimal set of vectors $\vec{m}_x$ and $\vec{v}_y$ resulting from this procedure, one will recover the maximum value of $\mathcal{W}$ for that set of vectors even if $\left|\vec{z}_y\right|$ equals to $\left|W_y\right|$ for some $y$ so that the value of the optimal corresponding $\mu_y$ is ill-defined. The fact that $\left|\vec{z}_y\right| = \left|W_y\right|$ for a given $y$, which is relevant for self-testing, will be indicated by a direct check. Furthermore, $\left|\vec{z}_y\right| = \left|W_y\right|$ for a given $y$ and for an optimal set of vectors $\vec{m}_x$ and $\vec{v}_y$ cannot be the case if the maximal $\mathcal{W}$ is not invariant under changing the value of the corresponding $\mu_y$.

We first notice that $W_2 = W_3 = 0$, so that $\mu_2 = \mu_3 = 0$ can be assumed for the search of the maximal $\mathcal{W}$. The case of $\mu_1 = 0$ is discussed in Section~\ref{sec:states} and leads to \eqref{eq:mmax} with \eqref{eq:u1max}-\eqref{eq:u4max} for $\vec{m}_x$. In this case,
\begin{align}
	|\vec{z}_2| = |\vec{z}_3| &= 4 \sqrt{ \frac{\cos^2\theta}{c_1^2+4\cos^2\theta} + \frac{\sin^2\theta}{c_2^2+4\sin^2\theta} }\nonumber\\
 &= \frac{8}{\sqrt{2(c_1^2+c_2^2+4)}} > 0 ,
\end{align}
so that $|\vec{z}_2| \neq |W_2|$ and $|\vec{z}_3| \neq |W_3|$.

Instead of checking the relation between $|\vec{z}_1|$ and $|W_1|$ for the optimal vectors corresponding to $\mu_1 = 0$, we consider the alternative option $\mu_1 = \sgn W_1$ for $W_1 \neq 0$ and $|\mu_1| = 1$ for $W_1 = 0$ (which is the case for $c_1 = c_2$) while keeping $\mu_2 = \mu_3 = 0$. For assessing this option, we return to \eqref{eq:W_general_transformed2}, which then takes the form
\begin{align}
	\mathcal{W} =& \left| \sum_x w_{x1} \right| \nonumber \\
    &+ \vec{v}_2 \cdot \left( \sum_x w_{x2} \vec{m}_x \right) + \vec{v}_3 \cdot \left( \sum_x w_{x3} \vec{m}_x \right) \nonumber \\
    =& \left| \sum_x w_{x1} \right| + \sum_x \vec{m}_x \cdot \left( w_{x2} \vec{v}_2 + w_{x3} \vec{v}_3 \right) .
\end{align}
For the optimal choice of $\vec{m}_x$, given by \eqref{eq:mmax}, this simplifies to
\begin{align}
	\mathcal{W} &= \left| \sum_x w_{x1} \right| + \sum_x \left| w_{x2} \vec{v}_2 + w_{x3} \vec{v}_3 \right| \nonumber \\
    &= 2 |c_1-c_2| + 2 \left| \vec{v}_2 + \vec{v}_3 \right| + 2 \left| \vec{v}_2 - \vec{v}_3 \right| .
\end{align}
It is easy to see that this expression is maximised in $\vec{v}_2$ and $\vec{v}_3$ by choosing these unit vectors to be orthogonal; in this case,
\begin{equation}\label{eq:Qp}
    \mathcal{W} = 2 |c_1-c_2| + 4\sqrt{2} =: Q' .
\end{equation}
is obtained.

For general values of $c_1$ and $c_2$, $Q'$ can be greater than $Q$ in \eqref{eq:Q}, which means that a degenerate measurement as Bob's first one can in fact be optimal. However, if we do not allow for arbitrary $c_1$ and $c_2$ in \eqref{eq:Qp} but substitute \eqref{eq:c1_of_B}-\eqref{eq:c2_of_B} linking them to the parameter $B$ of a semi-SIC POVM, the only solution to $Q = Q'$ in terms of $B$ will be $B = 1/16$, lying just outside the domain $B \in (1/16,1/12]$. If we add that
\begin{equation}
    Q(B=1/12) = 4 \sqrt{3} > Q'(B=1/12) = 4 \sqrt{2} ,
\end{equation}
we see that $Q > Q'$ for all $B \in (1/16,1/12]$. We thus conclude that $\mu_1 = 0$ is indeed the optimal choice for maximising $\mathcal{W}$ whenever the witness matrix $w$ is chosen for the purpose of self-testing a semi-SIC POVM.

\begin{acknowledgments}
We would like to thank Mateus Araújo, Sébastien Designolle, Miguel Navascués and Armin Tavakoli for their valuable input and enlightening discussions. We acknowledge the support of the EU (QuantERA eDICT) and the National Research, Development and Innovation Office NKFIH (No. 2019-2.1.7-ERA-NET-2020-00003).
\end{acknowledgments}

%

\end{document}